\documentclass[a4paper,10pt]{article}
\usepackage{epsfig}

\begin{document}

\title{\center{The re-analysis of the 1700 MeV structure of the P$_{11}$ partial wave using the \mbox{$\pi$N 
$\rightarrow$ $K\Lambda$} production data}}

\author{Sa\v sa Ceci, Alfred \v{S}varc\footnote{\emph{E-mail address:} Alfred.Svarc@irb.hr} \hspace*{0.25cm}and Branimir Zauner\\ \emph{Rudjer Bo\v{s}kovi\'{c} Institute}\\ \emph{Bijeni\v{c}ka c. 54}\\ \emph{10 000 Zagreb, Croatia}}
\maketitle
\begin{abstract}
{ \small We have used the Breit-Wigner resonance model with S$_{11}$, P$_{11}$ and P$_{13}$ resonances in the 
s-channel to re-analyze the old \mbox{$\pi$N $\rightarrow$ $K\Lambda$} data with the aim to establish the origin of 
the prominent structure in the total cross section in the vicinity of 1700 MeV. 
We have found the new set of resonance parameters enforcing the experimentally observed structure of the total cross 
section data simultaneously with the linear dependence of the differential cross sections with the cos$\, \theta$ in 
the energy range 1650~MeV~$<$ W $<$~1800~MeV. Due to the differential cross section linearity, the P$_{13}$ partial 
wave has been strongly attenuated in this model, and the total cross section structure is attributed to the resonant 
behavior of the P$_{11}$ partial wave. In this paper we show that, at least in the Breit-Wigner resonance model, it is 
not possible to achieve the detailed reproduction of the narrow 1700 MeV total cross section peak using the standard 
partial widths. To understand the phenomenon, a much narrower width of a resonant state, the N(1710)~P$_{11}$ in our 
case, is required ($\Gamma \approx 68$~MeV), but then the agreement of the model predictions with total cross section 
data at higher energies is lost. One way out is to allow for the existence of the second P$_{11}$ resonance in that 
energy range. The same feature is shown by the polarization data: the introduction of a much narrower resonance spoils 
the level of agreement which the Breit-Wigner resonance model is able to achieve with experiment, but the consistency 
is restored when a second resonance is introduced. Analyzing the $qqq$ or $qqqq\bar{q}$ nature of the recommended 
narrow P$_{11}$ structure in the neighbourhood of 1700~MeV we re-open (remind of) the possibility that another 
P$_{11}$ resonant state exists in addition to the standard N(1710)~P$_{11}$ PDG-resonance, and that one of the two 
states can be identified with the yet undiscovered cryptoexotic pentaquark state. To clarify the situation, we 
strongly recommend a re-measurement of the \mbox{$\pi$N $\rightarrow$ $K\Lambda$} process in the energy range 1650 MeV 
$<$ W $<$ 1800 MeV.}
\end{abstract}


\section{Introduction}
The recently undertaken extensive experimental searches for baryon states that can not be constructed with triplets of 
u, d and s quarks (exotic states) seem to have revealed significant positive results \cite{Kle03}. Consequently, the 
models predicting baryon states constructed out of more complicated quark configurations have been brought into focus.

The most popular recent models are the chiral soliton \cite{Dia97,Dia03} and $qqqq \bar{q}$ strong color-spin 
correlated models \cite{Jaf03}. Both models have fixed the scale by using the mass of the non-strange member of the 
pentaquark anti-decuplet, and octet $\otimes$ anti-decuplet configurations as the input. The first, chiral soliton 
model \cite{Dia97}, has assumed that the N(1710)~P$_{11}$ resonance exists, and has used its the mass as a scaling 
value for their non-strange pentaquark state. After the discovery of the $\Theta^+$ exotic state, the authors have 
improved the original model. In the original paper by Diakonov et.al. \cite{Dia97} the PDG N(1710)~P$_{11}$ has been 
directly identified as the member of the pentaquark anti-decuplet, and the model predictions for the partial width of 
that state have been compared with the experimentally established (PDG) values. In spite of the achieved reasonable 
agreement, the authors of ref. \cite{Dia97} have left the possibility of simultaneous existence of two nearby 1/2$^+$ 
resonances open, and justified it by referring to the findings of a possible degeneration of the N(1710)~P$_{11}$ 
state made in the coupled channel multi-resonance PWA done by Zagreb group \cite{PDG,Bat98}. 

As the exotic pentaquark states seemed to be confirmed, and
the original identification of the cryptoexotic 1/2$^+$ state with the N(1710)~P$_{11}$ resonance does not seem to 
pass the theoretical scrutinization~\cite{Pra04,Pak04}, the search should be continued to find a signal for the new 
1/2$^+$ anti-decuplet resonance corresponding either to the N(1646) of ref.~\cite{Dia03} (which in that model mixes 
with the N(1440), $\Lambda$(1600), $\Sigma$(1660) and $\Xi$(1690) octet) or to the N${_s}$($\approx$1700) of 
ref.~\cite{Jaf03}. In both cases the standard N(1710) P$_{11}$ PDG-resonance belongs to a pure N(1710), 
$\Lambda$(1810), $\Sigma$(1880) and $\Xi$(1950) octet, and exhibits no mixing with the pentaquark anti-decuplet.
On the other hand, the $qqqq \bar{q}$ strong color-spin correlated model \cite{Jaf03} has fixed its scale by 
identifying the the first - (${\bf [ud]^2 \bar{s}}$) state as $\Theta^+$, the second - (${\bf [ud]^2 \bar{d}}$) state 
as N(1440)-Roper resonance, and the N(1710)~P$_{11}$ state is identified with the third - (${\bf [ud] [su]_+ 
\bar{s}}$) hidden strangeness state N${_s}$.

Contrary to the fact that the exotic states can be unambiguously experimentally confirmed, the distinction between any 
three quark baryon state and a non-strange pentaquark configuration is a completely open, model dependent problem.
Namely, the pentaquarks, $\theta^+$(1540) and $\Xi^{--}$(1862) consist of solely of a $qqqq\bar{q}$ wave function 
because a \mbox{$qqq$ - gluon} wave function does not produce baryons with their quantum numbers.
As we may write down the Fock space expansion of any/each baryon as a linear combination of $qqq$ and $qqqq\bar{q}$ 
states, we are left with a problem of defining a criteria to distinguish between a three quark state with the diquark 
contribution coming from the sea quarks, and the genuine pentaquark states. It can be done only by comparing the 
experimentally established characteristics of the resonant state with the corresponding values coming from the $qqq$ 
and $qqqq\bar{q}$ model predictions, and defining some selection rules \cite{Lan99}. It is completely clear that we 
can only give a positive hint that a state is a pentaquark state, and not a definite proof.

The analysis of the P$_{11}$ resonant structure in the whole energy range, consequently, becomes the issue of prime 
importance. This study is motivated by the search for a non-strange partner of the exotic pentaquark $\Theta^+$ state. 
Since the N(1710)~P$_{11}$ state is one of such candidates its properties are of prime interest. Let us point out that 
the extraction of the N(1710)~P$_{11}$ resonance parameters from experiment is very important even if the existence of 
$\Theta^+$ state would not be confirmed in future because of variety of other issues related to the analytic structure 
of pion-nucleon amplitudes.

 We shall approach this problem by trying to understand the origin of the prominent, and quite narrow peak in the 
\mbox{$\pi$N $\rightarrow$ $K\Lambda$} total cross section data which is not yet fully understood. Simple Breit-Wigner 
resonance models using the PDG parameters \cite{PDG} and the more sophisticated coupled-channel models 
\cite{Wen04,Pen02,Shk05} do give certain level of peaking around 1700 MeV, but far from enough as indicated by data. 
In this paper we give a mechanism how the structure of the data (strong peaking of the total cross section data of 
$\sigma _{tot} \approx$ 900 mb) can be accounted for in details without disturbing measured observables elsewhere. We 
show that it is necessary to increase the branching ratio of the P$_{11}$ resonance to the $K\Lambda$ channel 
simultaneously with reducing the N(1710)~P$_{11}$ total width, but then the agreement of the model predicted total 
cross section at energies of W $\approx$ 1750 MeV is significantly spoiled. One way out is to allow for the existence 
of the second resonant state which we identify to be as well in the P$_{11}$ partial wave. 
The functioning of the mechanism is demonstrated in a simple Breit-Wigner resonance model, but we expect it to work 
when extended to a coupled channel calculation. 

In our simple Breit-Wigner resonance model we are focused to the cross section data only, but as a comparison we show 
the model predictions for the longitudinal polarization data as well. That aspect of our publication requires an 
additional discussion. As it is well known, even the elastic channel polarization data are presently hard to be 
understood, and the need still exists to create an "improved program \' a la KH80 which also gives good results for 
spin rotation data above 1 GeV and large angles" \cite{Hoh05}. So it is no miracle that our simple Breit-Wigner 
resonance model quite poorly reproduces the longitudinal polarization $K\Lambda$ data. It better be so, because 
otherwise no one would believe that our calculation is correct. This paper relies upon the belief that something else, 
and not reproducing the data is important: the same pattern of behavior, seen in cross section data, is followed for 
polarization as well. Namely, i) model based on the standard type of P$_{11}$ resonance (branching ratio $\approx 15$ 
- 30 \%, partial width $\approx$ 100 - 200 MeV) only tolerably reproduces the data; ii) the modification (reduction) 
of $K\Lambda$ partial width required for reproducing the 1700 MeV peak in the total cross section spoils the level of 
agreement of the Breit-Wigner resonance model with experiment; iii) the addition of a second resonance in that energy 
range restores the agreement of model predictions for longitudinal polarization to the initially achieved level. Our 
intention in case of polarization is not to discuss the level of consistency with experiment, but the dynamics of the 
procedure: if one wants to improve on one part, something else is spoiled, so a new mechanism (a new resonance in our 
case) is needed to repair the damage. 

This paper contains: the detailed explanation of the prominent, and yet not fully understood total cross section peak 
in the \mbox{$\pi$N $\rightarrow$ $K\Lambda$} process \cite{Wen04,Pen02,Shk05} in terms of sometimes questioned 
N(1710)~P$_{11}$ resonance \cite{Arn85,Arn95,Arn04} and another, not yet discussed P$_{11}$ resonant state; the 
discussion of the proper interpretation of these states upon its acceptance; and finally advocating new experiments in 
order to improve the experimental situation in the \mbox{$\pi$N $\rightarrow$ $K\Lambda$} channel which is the most 
sensitive process for analyzing the N(1710)~P$_{11}$ resonance. The problem of the width of the P$_{11}$ resonant 
state will be addressed in particular.
As the width of the physical N(1710)~P$_{11}$ state is fairly undetermined ($90<\bar{\Gamma}_{\rm PDG}< 480$~MeV) - 
\cite{PDG} and quark models predict different ranges for the P$_{11}$ width (of the order of 200~MeV for the standard 
three quark models, and of the order of 40~MeV for the pentaquark models), we wonder if it is possible that the 
observed physical state is the admixture of both, three quark and a pentaquark states.

An attempt of finding the new 1/2$^+$ state has been done in ref. \cite{Arn04a}, and was based on admixing an 
additional narrow 1/2$^+$ resonance to the GWU PWA of ref. \cite{Arn85,Arn95,Arn04} in a way to make a standard PWA 
sensitive to narrow resonances. It has revealed two possible candidates for a new 1/2$^+$ state at 1680 and 1730 MeV.

\section{The \mbox{$\pi$N $\rightarrow$ $K\Lambda$} data and the existence of the N(1710)~P$_{11}$ resonant structure}
Without the $K\Lambda$ data even the sole existence of N(1710)~P$_{11}$ resonance is seriously questioned in the 
energy-dependent coupled-channel Chew-Mandelstam K matrix analysis \cite{Arn85,Arn95,Arn04}. However, due to the 
constraints coming from the crossed channels, the fully analytic Karlsruhe-Helsinki PWA \cite{Hoh83} undoubtedly sees 
it, and qualifies it as being strongly inelastic. Very little doubt about its existence is left when the inelastic 
data are included in the analysis. The coupled-channel T-matrix Carnegie-Mellon Berkeley (CMB) type models report it 
and claim that it is strongly inelastic \cite{Bat98,Cut79,Vra00}. The coupled channel K-matrix analyses (University of 
Giessen) agrees with the existence of the N(1710) P$_{11}$ state, but the status of its importance for the 
\mbox{$\pi$N $\rightarrow$ $K\Lambda$} process varies from important \cite{Pen02,Shk05} to dominant \cite{Feu98}. 
Another coupled channel K-matrix analyses \cite{Wen04} also reports the N(1710) P$_{11}$ state as one of the important 
contributions, but does not offer the separation into individual resonance contributions so we have no information 
about its relevance in that model.
 
Putting aside the fact that the problem of identifying resonances seems to be more of fundamental then of technical 
nature\footnote{Namely, in the paper by Cutkosky and Wang \cite{Cut90} it has been shown that the coupled-channel 
method predicts the existence of additional N(1710)~P$_{11}$ pole in the energy range where the energy-dependent 
coupled-channel Chew-Mandelstam K matrix method does not see anything, when fitting the identical set of single energy 
T-matrices. It has been explicitly 
concluded that the differences in pole structure in the two afore described models arise from the different 
parameterization of the energy dependence, rather than differences in the data. The answer to this puzzle is yet to be 
given, but it is conceivable that the used number of channels is insufficient, and that more rigorous inclusion of 
$K\Lambda$ channel is needed.} it is important to stress that the very strong peak, formed when the $\approx$ 1700 MeV 
data points from refs. \cite{Bak77,Bak78,Dyc69,Jon71} are taken seriously, is not reproduced in any of these models. 
All models do see the peaking structure, but is definitely too low. No attempts have been made in these models to give 
a physical explanation about what is the possible physical origin of a serious underestimation of the peak value 
($\approx$ 70 \%). Our intention is to show how it can be accounted for in a simple Breit-Wigner resonance model.

The experimental data sets for the process \mbox{$\pi$N $\rightarrow$ $K\Lambda$} are available for quite some time 
\cite{Bak77,Bak78,Dyc69,Jon71,Kna75}. The total cross section data clearly show a distinct structural behavior in the 
vicinity of 1700 MeV and the differential cross sections \cite{Bak77,Bak78,Kna75} show a clearly recognizable 
linearity in cos$\,\theta$ up to 1850 MeV. As the experiments admit the systematic error of 8 - 15 \% in absolute 
normalization, the observed structure is smeared in energy making any conclusions about the profile of the structure 
very difficult. Even when the agreement about the position and the width of the structure is achieved, the 
interpretation of its origin remains unclear. It can be interpreted either as a genuine T-matrix pole, namely the 
signal of a resonance, or as a cusp effect resulting from the opening of the $K\Sigma$ channel. The distinction 
between the resonant and the cusp effect interpretations is as well non-trivial. To claim that the structure is a 
genuine T-matrix pole requires a full scale coupled-channel analysis which manifestly incorporates the appearance of 
cusp effects, and a clear and unambiguous search for the poles in the complex energy plane via well defined analytical 
continuation. Anything less (Argand diagrams, fits to different single-resonance models,..) fails on the basis of 
first principles as it has been demonstrated in the search for the dibaryon resonance hidden in the $^1$D$_2$ and 
$^3$F$_3$ partial waves in the pp $\rightarrow$ pp elastic scattering \cite{Loc86}.
Partly because of data dissipation, and partly because of technical complexity, the \mbox{$\pi$N $\rightarrow$ 
$K\Lambda$} process has not yet been included as the part of the data base in the existing coupled-channel 
$\pi$-nucleon partial wave analyses (PWA) \cite{Bat98,Arn85,Arn95,Arn04,Hoh83,Cut79,Vra00,Man84,Man92,Gre99} until 
late 90-es \cite{Feu98}, and even then, according to the authors themselves, play a minor role because of large errors 
and are included for completeness only.

The N(1710)~P$_{11}$ resonance parameters have been for the first time extracted from the \mbox{$\pi$N $\rightarrow$ 
$K\Lambda$} data in a single-channel energy-dependent phase shift analysis \cite{Bak77}, which was soon afterwards 
upgraded using the new set of data \cite{Bak78}. Because of using only one channel, the extracted resonances for the 
S$_{11}$, P$_{11}$ and P$_{13}$ partial waves do not necessarily describe all $\pi$-nucleon channels at the same time, 
so the single-channel result should be coordinated with the values obtained in the multi-channel PWA. That has been 
attempted in ref. \cite{Feu98}.

\section{Formalism}
In this article we only use a simple s-channel Breit-Wigner resonance model with no background and no t-channel 
contributions taken into consideration. This is a very strong simplification of reality, but we do believe that the 
main conclusions of the paper are of qualitative and not of quantitative nature, and that such a simple model 
suffices. For getting the proper quantitative answers the full coupled channel calculation will be undertaken in near 
future. 
 
The differential cross section and longitudinal polarization for the \mbox{$\pi$N $\rightarrow$ $K\Lambda$} process 
are standardly described in terms spin-flip and non-spin-flip amplitudes \cite{Hoh83}:
\begin{eqnarray} 
 \frac{{\rm d} \sigma}{{\rm d}\Omega} (s, z)& = &  \ |G(s, z)|^2 + |H(s, z)|^2 \  \\
			P(s, z) &=& \frac{2 \ {\rm Im} \ [ G(s, z) H^*(s, z)]}{\ |G(s, z)|^2 + |H(s, z)|^2 } \ . 
\end{eqnarray}
Their expansions in terms of partial wave amplitudes $T_{l^\pm}$ are given as:
\begin{eqnarray} 
 G (s, z) & = & \frac{1}{q} \ \sum_{l=0}^{\infty} \ [ \ (l+1)T_{2I,2l^+}^{\pi N,K\Lambda}(s) + l T_{2I,2l^-}^{\pi 
N,K\Lambda}(s) \ ] \ P_l(z); \nonumber \\
	 H (s, z) & = & \frac{1}{q} \ \sum_{l=1}^{\infty} \ [ \ T_{2I,2l^+}^{\pi N,K\Lambda}(s) - T_{2I,2l^-}^{\pi 
N,K\Lambda}(s) \ ] \ \sin \theta \ P_l^{'}(z), \nonumber \\	
z &=& \cos \theta,	
\end{eqnarray}
where $ 2l_\pm$ indicates that the total angular momentum is $2J = 2l \pm 1$, s is a Mandelstam variable and  $q$ is a 
center of mass momentum. The $P_l(z)$ are Legendre polynomials and $P_l^{'}(z)$ are their derivatives. \\
The $T_{2I,2l^ \pm}^{\pi N,K\Lambda}(s)$ is given as:
\begin{eqnarray} 
T_{2I,2J}^{\pi N,K\Lambda}(s)&=& \sqrt{T_{2I,2J}^{\pi N} (s) \ \cdot \  
T_{2I,2J}^{K\Lambda}(s) }, \nonumber
\end{eqnarray}
with elastic channel T-matrices given as Breit-Wigners resonances: 
\begin{eqnarray} 
T_{2I,2J}^{ch}(s)&=&\frac{ \frac{\Gamma _{ch} (s)}{2}}{M- \sqrt{s} -i\frac{\Gamma (s)}{2}}, \\ 
\Gamma (s) &=& \sum _{ch} \Gamma _{ch}(s),  \nonumber 
\label{eq:sres}
\end{eqnarray}
where we sum over channels $ch$ = $\pi N$, $K \Lambda$.  \\
$M$, $\Gamma(s)$ and $\Gamma _{ch}(s)$ are resonance mass, total and partial widths. \\
The energy dependence of the Breit-Wigner width is taken over from \cite{Bat98}, and is given as: 
      \begin{equation}
          \Gamma _{ch} (s) = \Gamma _{ch}(M^2) \times
          \left\{
          \begin{array} {cl}
               \left( \frac{q_{ch}(s)}{q_{0ch}} \right)^{2L+1}
                & $for$ \  q_{ch}(s)  < q_{0ch}          \vspace{3mm} \\
             \left( \frac{2 \ q_{ch}(s) }{q_{ch}(s)  + q_{0ch}} \right)^{2L+1}
                & $for \ $ q_{ch}(s)  > q_{0ch}
          \end{array}
          \right.
      \end{equation}
    where $q_{ch}$(s) is the momentum of the channel meson.
	      \begin{equation}
          q_{ch} (s) =
         \frac{\sqrt{\left(s-(M_{c}+m_c)^2\right)
                     \left(s-(M_{c}-m_c)^2\right)}} {2 \sqrt{s}}; \ \ \ q_{0ch} \stackrel{def}{=} q_{ch}(M^2).
      \end{equation}
      with $M_{c}$ = $m_N$, $m_K$ and $m_c$ = $m_\pi$, $m_\Lambda$. \\
The t-channel resonant contribution of $K^*$(892) meson is in this calculation neglected, because the product of the 
$K^* \pi K $ and $K^{*}\Lambda N$ coupling constants, which enters the model, is poorly determined, and due to the 
kinematic behavior of the $K^*$(892) propagator influences the shape of the differential cross section significantly 
only at energies far above the domain of interest of 1650-1750 MeV \cite{Feu98}. The square root recipe for the 
$T_{2I,2J}^{\pi N,K\Lambda}$ matrix is generally valid in a single resonance approximation, but is believed to be a 
fair approximation within a resonance energy domain for any full calculation. Therefore, the use of a Breit-Wigner 
resonance model should be sufficient to establish the relevant resonance parameters near the top of the resonance.

The collection of data sets which we have used for a comparison with the predictions of our model 
\cite{Bak77,Bak78,Dyc69,Jon71,Kna75} declare an 8 - 15 \% systematic error in the absolute normalization, therefore an 
overall normalization of absolute scale is in order. The smearing of the structure in the total cross section, due to 
the normalization uncertainties in the reported experiments has been lessened by creating the amalgamated data set. 
Because three out of four experiments \cite{Bak77,Bak78,Dyc69,Kna75} report the peak of 920 $\mu$b at the energy of 
1694~MeV, we have chosen to normalize all data accordingly. In addition, we have shifted the whole Knasel et al. data 
set \cite{Kna75} up in energy for 9 MeV. That step is open for criticism because it implicitly questions the energy 
calibration of that particular experiment. That is not our intention. At this point, we are primarily interested in 
the shape of the structure, namely its width, because we tend to interpret it as a possible narrow non-strange 
pentaquark candidate. The mass of the resonance is not of our prime concern. By shifting the peak of the structure to 
the same energy we automatically extract the common width from all experiments, because otherwise the width is smeared 
and comes out bigger then actually extracted from each experiment separately. We could have shifted other three 
experiments 9 MeV down in energy and obtain the same width, but lower mass.

\section{Results and conclusions}
\begin{figure}[!h]
\begin{center}
\epsfig{figure=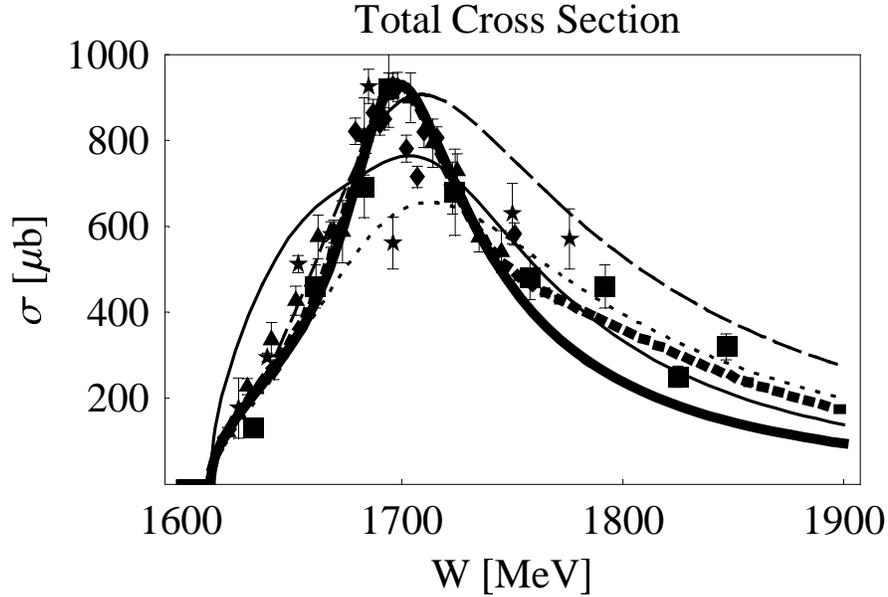,height=8cm,angle=0}
\caption{The agreement of the amalgamated experimental data for the total cross section (ref. 
\protect\cite{Dyc69}-stars; ref. \protect\cite{Jon71}-triangles; ref. \protect\cite{Kna75}-diamond and ref. 
\protect\cite{Bak77,Bak78} boxes) with the Breit-Wigner resonance model predictions using different inputs for the 
resonance parameters: "standard" (PDG) set (thin solid line); $Sol \ 1$ (dotted line); $Sol \ 2$ (dashed line), $Sol \ 
3$ (thick solid line) and $Sol \ 4$ (thick dotted line).}
\label{fig:1}
\end{center}
\end{figure}

\begin{figure}[!hbt]
\begin{center}
\epsfig{figure=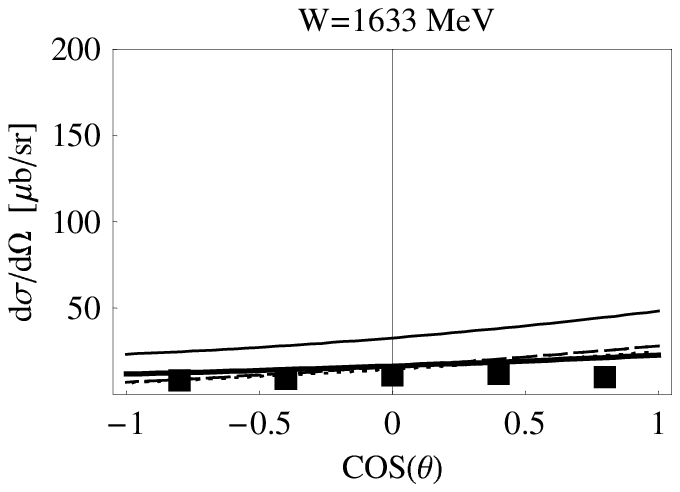,width=3.8cm,angle=0} \epsfig{figure=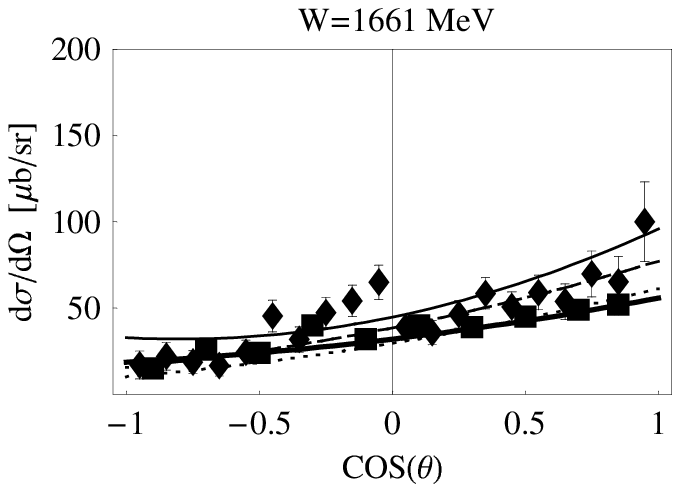,width=3.8cm,angle=0} 
\epsfig{figure=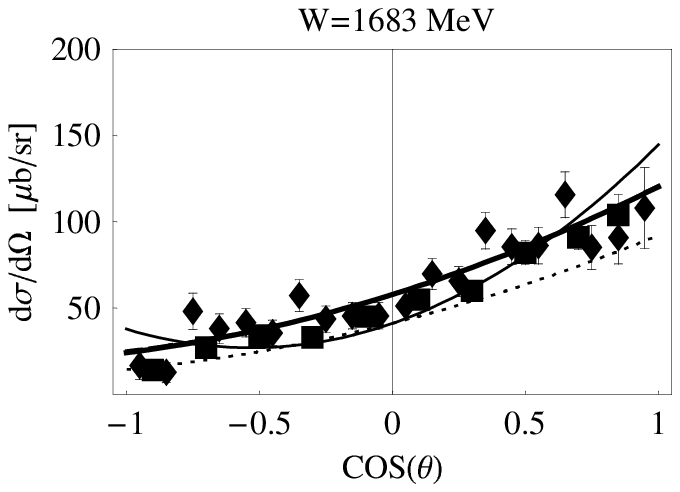,width=3.8cm,angle=0} \\
\epsfig{figure=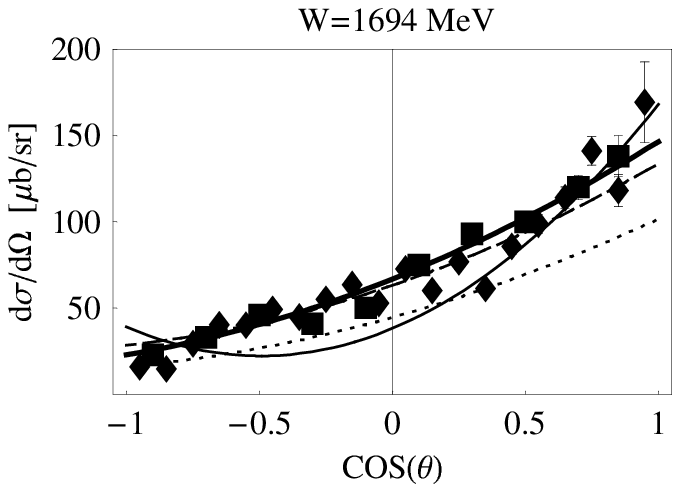,width=3.8cm,angle=0} \epsfig{figure=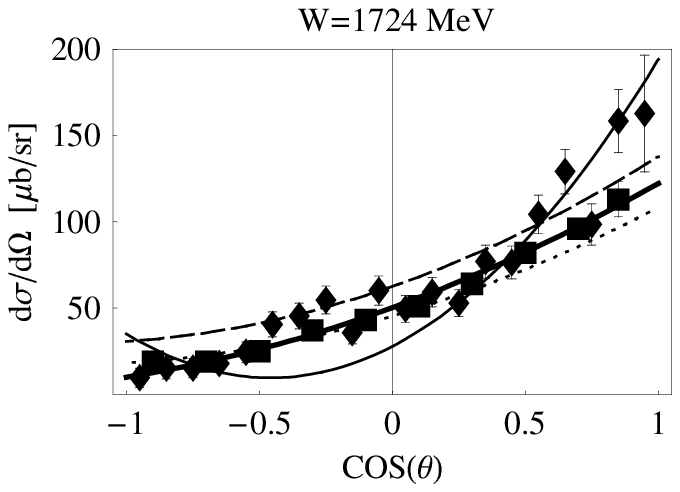,width=3.8cm,angle=0} 
\epsfig{figure=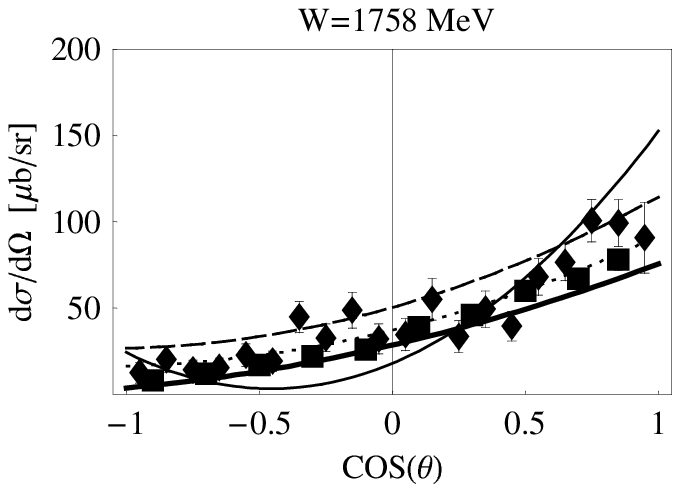,width=3.8cm,angle=0} \\
\epsfig{figure=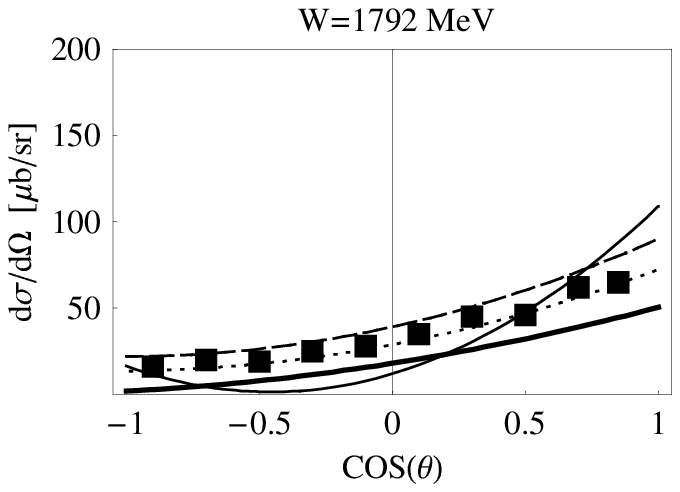,width=3.8cm,angle=0} \epsfig{figure=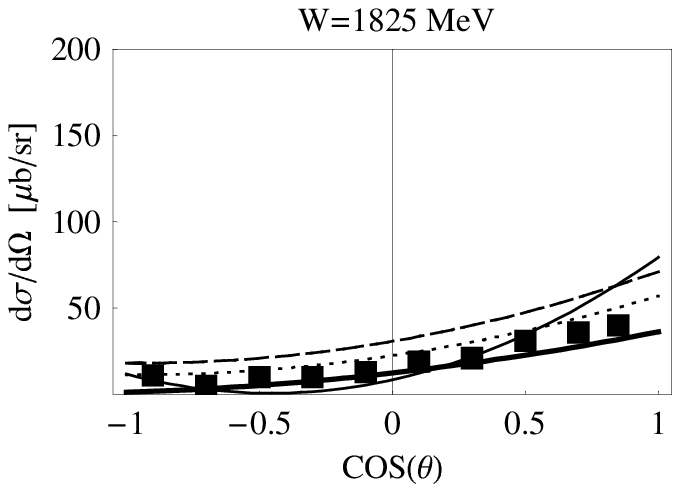,width=3.8cm,angle=0} 
\epsfig{figure=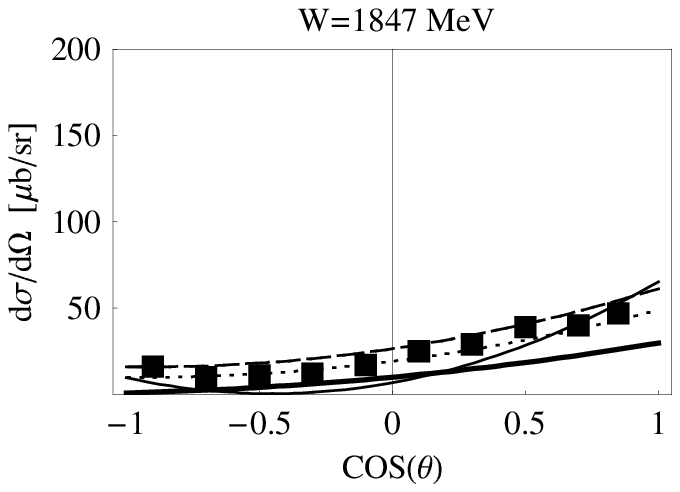,width=3.8cm,angle=0} 
\caption{The agreement of the amalgamated experimental data for the differential cross section (ref. 
\protect\cite{Kna75} - diamonds and ref. \protect\cite{Bak78} - boxes); with the Breit-Wigner resonance model 
predictions using different inputs for the resonance parameters: "standard" (PDG) set (thin solid line); $Sol \ 1$ 
(dotted line); $Sol \ 2$ (dashed line) and $Sol \ 3$ (thick solid line).}
\label{fig:2}
\end{center}
\end{figure}

\begin{figure}[!hbt]
\begin{center}
\epsfig{figure=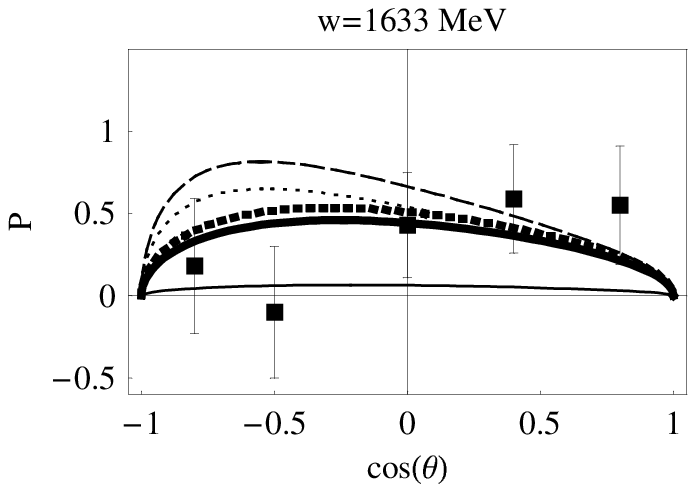,width=3.8cm,angle=0} \epsfig{figure=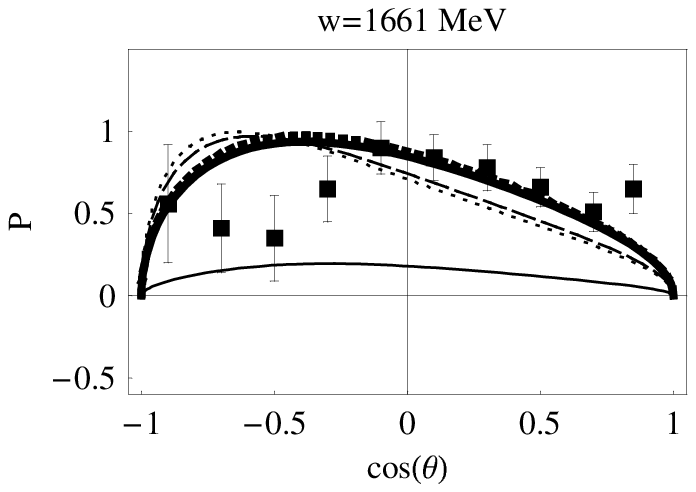,width=3.8cm,angle=0} 
\epsfig{figure=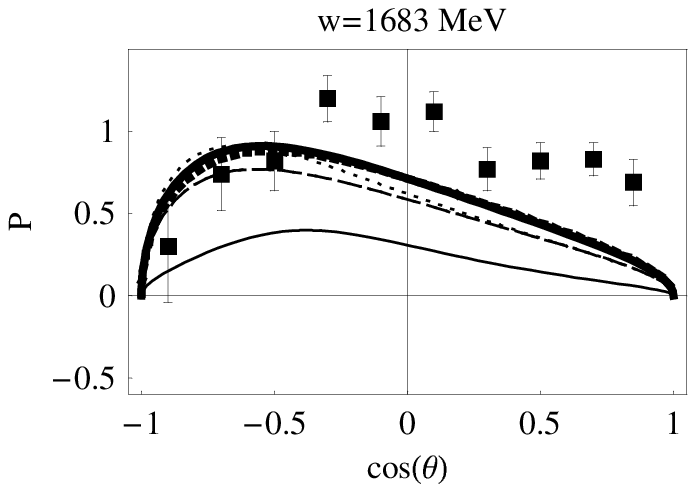,width=3.8cm,angle=0} \\
\epsfig{figure=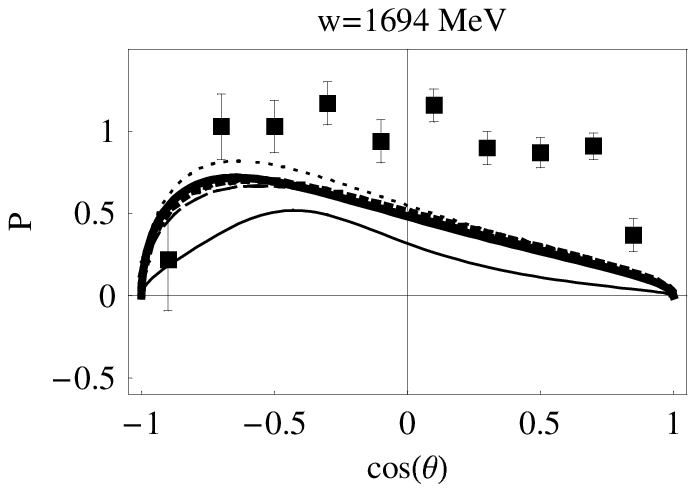,width=3.8cm,angle=0} \epsfig{figure=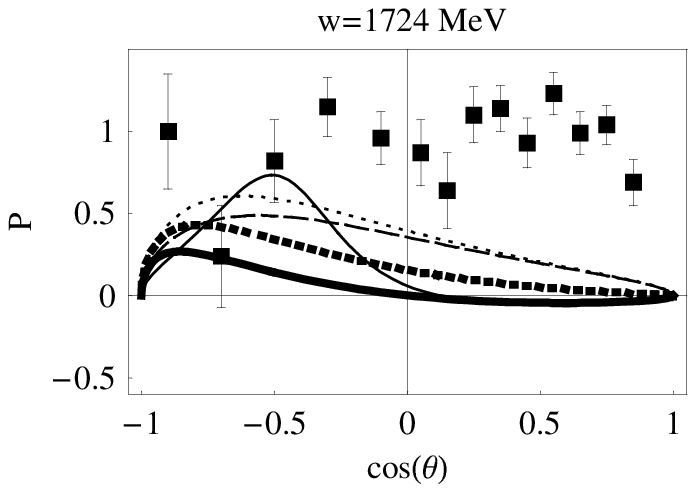,width=3.8cm,angle=0} 
\epsfig{figure=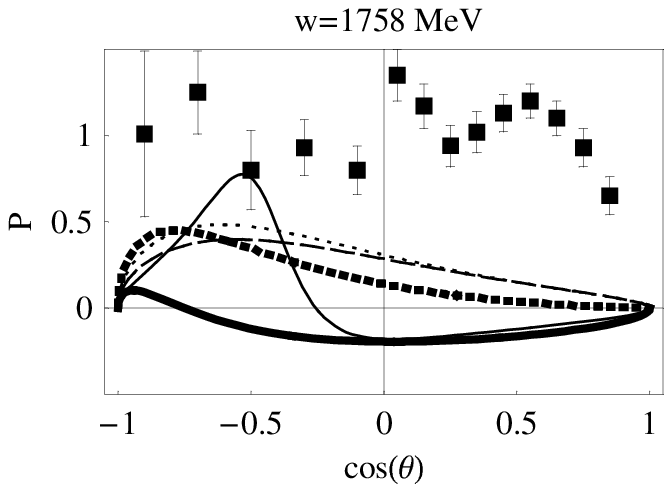,width=3.8cm,angle=0} \\
\epsfig{figure=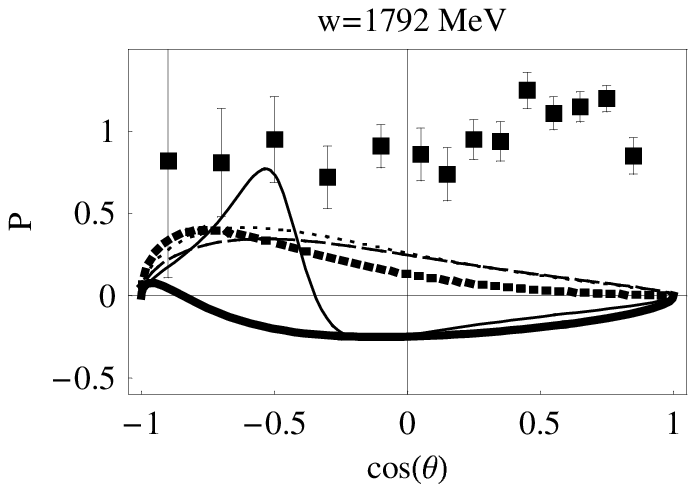,width=3.8cm,angle=0} \epsfig{figure=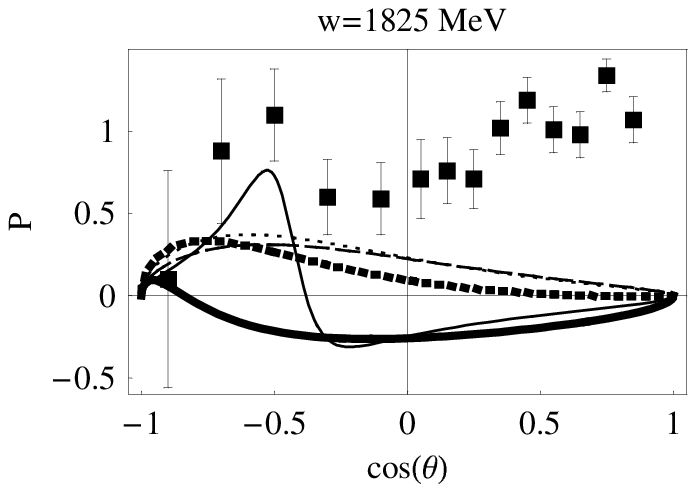,width=3.8cm,angle=0} 
\epsfig{figure=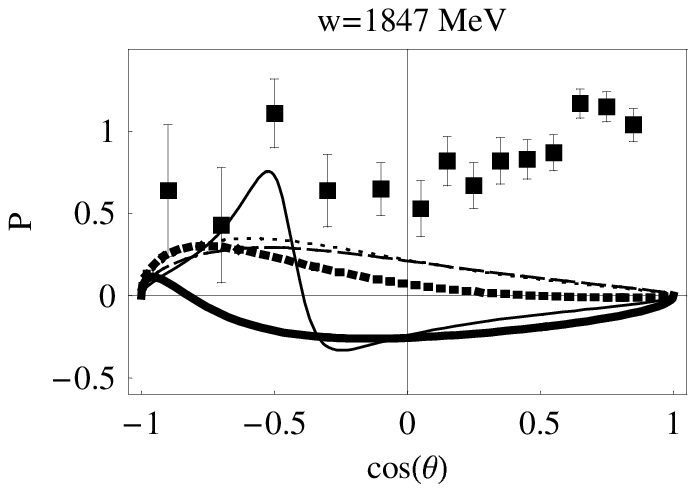,width=3.8cm,angle=0} 
\caption{The agreement of the polarization data (ref. \protect\cite{Bak78} - boxes); with the Breit-Wigner resonance 
model predictions using different inputs for the resonance parameters: "standard" (PDG) set (thin solid line); $Sol \ 
1$ (dotted line); $Sol \ 2$ (dashed line),$Sol \ 3$ (thick solid line) and $Sol \ 4$ (thick dotted line).}
\label{fig:3}
\end{center}
\end{figure}

The Breit-Wigner resonance model calculation has been repeated using the standard set of parameters for the 
N(1650)~S$_{11}$, N(1710)~P$_{11}$ and N(1720)~P$_{13}$ resonances of ref. \cite{PDG}, the parameters are given in 
table 1 denoted as "PDG", and the agreement with experiment is given in figures \ref{fig:1} through \ref{fig:3} as 
thin solid line. 
That choice of parameters gives the overall reproduction of the absolute value of the total cross section, and manages 
to reproduce the shape of the angular distribution only at W=1683 MeV. However, it fails miserably in reproducing the 
$\sigma _{tot} \approx$ 900 nb peak and the shape of differential cross section data at other energies. The agreement 
for the polarization data is, as expected for the Breit-Wigner resonance model, poor but qualitatively correct. That 
should not come as a surprise because the Breit-Wigner resonance model is only a crude approximation of reality 
because the polarization, being an amplitude interference effect, is very dependent on the $G$/$H$ relative phase. The 
experimental data are tolerably reproduced at energies lower then 1700 MeV, but the disagreement is strong for the 
forward angles, and especially at higher energies. 

Being dissatisfied with the predictions of the Breit-Wigner resonance model with PDG resonance parameters we have 
repeated the fit in order to obtain a good description of the absolute value of the total cross section maintaining 
the shape of the differential cross sections of the \mbox{$\pi$N $\rightarrow$ $K\Lambda$} (linear in cos$\, \theta$ 
as indicated by experimental data of ref. \cite{Bak77,Bak78,Kna75})\footnote{This has technically been done by 
dividing the total $\chi ^2$ into two equally contributing sub parts: one originating from the total, and second from 
the differential cross sections.}

We have fitted the amalgamated data set fixing the masses and $\pi$N partial widths of the N(1650)~S$_{11}$, 
N(1710)~P$_{11}$ and N(1720)~P$_{13}$ resonances to the values given in ref. \cite{Bat98}, and varying only the 
$K\Lambda$ branching ratios in the Breit-Wigner parameterization given in equation (\ref{eq:sres}). The energy 
dependence of the Breit-Wigner resonance width, which ensures the correct threshold and high energy behavior is for 
all solutions kept as in ref. \cite{Bat98}.
\begin{center}
\begin{table}[!htb]
\caption{Resonance parameters for the Breit-Wigner resonance model.}
\begin{tabular}{@{}c c c c | c @{ }c @{ }c | c @{ } c@{ } c | c@{ } c@{ } c}
\hline \hline
\multicolumn{13}{c}{ }\\ [-1.5ex]
{} & \multicolumn{3}{c}{$M[MeV]$} & \multicolumn{3}{c}{$\Gamma[MeV]$} & \multicolumn{3}{c}{$x_{\pi N}[\%]$} &
\multicolumn{3}{c}{$x_{K\Lambda}[\%]$} \\[+1ex]
\cline{2-13}
{}& {}& {}& {}& {}& {}& {}& {}& {}& {}& {}& {}& {}\\[-1.5ex]
{}&$S_{11}$&$P_{11}$&$P_{13}$&$S_{11}$&$P_{11}$&$P_{13}$&$S_{11}$&$P_{11}$&$P_{13}$&$S_{11}$&$P_{11}$&$P_{13}$\\ 
[+0.5ex]
\hline \hline
{}& {}& {}& {}& {}& {}& {}& {}& {}& {}& {}& {}& {} \\[-1.5ex]
$PDG$ & 1650 & 1710 & 1720 & 150 & 100 & 150 & 70 & 15 & 15 & 7 & 15 & 6.5 \\ 
{}& {}& {}& {}& {}& {}& {}& {}& {}& {}& {}& {}& {} \\[-2.5ex]
$Sol \ 1$ & 1652 & 1713 & 1720 & 202 & 180 & 244 & 79 & 22 & 18 & 2.4 & 23 & 0.16 \\ 
{}& {}& {}& {}& {}& {}& {}& {}& {}& {}& {}& {}& {} \\[-2.5ex]
$Sol \ 2$ & 1652 & 1713 & 1720 & 202 & 180 & 244 & 79 & 22 & 18 & 2.4 & 35 & 0.16 \\ 
{}& {}& {}& {}& {}& {}& {}& {}& {}& {}& {}& {}& {} \\[-2.5ex]
$Sol \ 3$ & 1652 & {\bf 1700} & 1720 & 202 & {\bf 68} & 244 & 79 & 22 & 18 & 3 & 32 & 0.16 \\ 
{}& {}& {}& {}& {}& {}& {}& {}& {}& {}& {}& {}& {} \\[-2.5ex]
$Sol \ 4$ & 1652 & {\bf 1700} & 1720 & 202 & {\bf 68} & 244 & 79 & 22 & 18 & 3 & 29 & 0.16 \\ 
		 & & {\bf 1775} & & & {\bf 154} & & & 28 & & & 3 & \\
\hline\hline
\end{tabular}
\label{table1} 
\end{table}
\end{center}

As a guidance for the fitting strategy we have used the fact that no existing model reproduces the $\sigma _{tot} 
(1700) \approx$ 900 nb. First we have required the reproduction of the peak value of the lower energy part of the 
total cross section. In that way we have obtained $Sol \ 1$ (dashed line in figures \ref{fig:1} through \ref{fig:3}). 
The $Sol \ 1$ offers the good reproduction of the lower energy part of the total cross data, but strongly over-shots 
the higher energy part. Aiming to reproduce the higher energy part of the total cross section we have obtained $Sol \ 
2$ (dotted line in figures \ref{fig:1} through \ref{fig:3}). The Breit-Wigner parameters of both fits are given in 
table 1. 

Notice that we do not offer any error bars for the resonance parameters. As the existing experimental data are of 
extremely poor quality and in mutual disagreement in the important 1700 Mev range, our analysis throughout this 
publication is of a qualitative nature only. We make no attempt for any uncertainty considerations of the fit because 
we believe that it is useless and misleading to discuss the level of confidence for fits of such a low quality data.
 \\ \noindent
The solutions $Sol \ 1$ and $Sol \ 2$ have the following features: 

As shown in figure \ref{fig:1} calculated \mbox{$\pi$N $\rightarrow$ $K\Lambda$} cross sections {\it can not} 
simultaneously describe the amalgamated data set throughout the whole energy range from 1650 to 1800~MeV for any 
choice of $K\Lambda$ branching fraction for the resonance masses and widths of ref. \cite{Bat98}. The $Sol \ 1$ fits 
the lower part of the energy range and overshoots the 1750/1800 MeV part, while $Sol \ 2$ undershoots the 
\mbox{W $\approx$ 1700 MeV} and fits the high energy part, very similar to the results of coupled-channel calculations 
given in ref. \cite{Wen04,Pen02,Shk05}. If the amalgamated data set gets the experimental confirmation the 
N(1710)~P$_{11}$ resonance can not be as wide as presently expected \cite{PDG}. 

The agreement of differential cross section data with predictions of our model is given in figure \ref{fig:2} with 
dashed lines for $Sol \ 1$ and dotted lines for $Sol \ 2$. The shape of the differential cross section is 
significantly improved throughout the whole energy range. The relative importance of the N(1720)~P$_{13}$ resonance in 
our model notably decreases. To account for the loss of flux, the contribution of the $P_{11}$ resonance to $K\Lambda$ 
channel significantly rises. The branching ratio of $S_{11}$ resonance to $K\Lambda$ channel turns out to be somewhat 
smaller then previously believed. 

The agreement of the longitudinal polarization with experimental data, given in figure \ref{fig:3}, is for both 
solutions improved with respect to the results with standard PDG solutions (full thin line), but still completely 
unsatisfactory; and that is typical for such a simple model. 

In order to improve the agreement of the Breit-Wigner resonance model with the amalgamated data set in the $\sigma 
_{tot} (1700) \approx$ 900 nb range, we have performed a fit where we have released the width of the N(1710)~P$_{11}$ 
resonance. We have obtained $Sol \ 3$, given in table 1, and compared with experimental data in figures \ref{fig:1} 
through \ref{fig:3} (thick solid line). We have obtained the best agreement with the experiment for a strongly 
inelastic but quite narrow $P_{11}$ resonance, $\Gamma = 68$ MeV.

The conclusion about the width of the N(1710)~P$_{11}$ resonance is valid only under the condition that its branching 
fraction to the $K \Sigma$ channel is small. Otherwise the cusp effect might effectively reduce the width obtained 
within the framework of the Breit-Wigner resonance model. However, according to the ref. \cite{PDG}, that condition 
seems to be accomplished.

However, our $Sol \ 3$ has two unwanted features: it has spoiled the agreement of the higher energy part of the total 
cross section together with the level of agreement of polarization predictions with experiment. The $Sol \ 3$ is 
definitely too low in total cross section predictions in the higher energy range ( W $\approx 1750$ MeV) and the 
agreement of the model prediction curves with the experimental values of the polarization data is spoiled in the sense 
that it is returned to the level of (dis)agreement that the PDG solution has. 

We conclude that only the reduction of the $P_{11}$ partial width from 180 MeV to 68 MeV, as obtained in $Sol \ 3$, is 
not enough to consistently explain the size of the total cross section peak of $\approx$ 900 nb together with the 
correct overall energy behavior. So, we need another contribution. As $P_{13}$ partial wave is attenuated in our 
model, we propose another resonance in the 1750 MeV range in the $P_{11}$ partial wave. 

We have used a simple K - matrix unitarized addition of an extra P$_{11}$ Breit-Wigner resonance\footnote{ 
$T_{tot} = \frac{K_{tot}}{1-i \ K_{tot}}; \ \ \ K_{tot}=\sum _{i=1}^{N} K_i; \ \ \ K_i= \frac{T_i}{1+i \ T_i}$} to 
form a new solution (now we have two nearby resonances in the P$_{11}$ partial wave as suggested in ref. 
\cite{PDG,Bat98}). It is interesting to observe that the new solution is consistent with the present quality of 
experimental data, and at the same time completely reproduces the $\sigma _{tot} \approx$ 900 nb peak together with 
the higher energy range behavior, without simultaneously destroying the existing "agreement" of the model predicted 
polarization data. 

The addition of a second resonance to the P$_{11}$ channel with parameters: M$_1$~=~ 1775~ MeV, $\Gamma _1$~=~154 MeV, 
x$_{\pi N}$~=~28 \% and x$_{K\Lambda}$ = 3 \%, given as $Sol \ 4$ in table \ref{table1}, reproduces the experimental 
total cross section data comparably to {\it Sol 3}\footnote{To compensate the increase of the total cross section the 
branching ratio of the first resonance is reduced from 32 \% to 29 \%.} - thick dotted line in figure \ref{fig:1}, and 
simultaneously restores the agreement of polarization predictions of $Sol \ 3$ to the originally achieved level - 
thick dotted line in figure \ref{fig:3}. If not reproduced correctly, the polarization predictions at least follow the 
general pattern that the agreement is improved when the second resonance is added. The change in differential cross 
section in figure \ref{fig:2} is in practice indistinguishable from values of $Sol \ 3$. 

It is clear that the addition of the second resonance in a K-matrix unitarized way does not affect the dominant 1710 
MeV peak of $Sol \ 3$, but raises the total cross section in the 1780 MeV range where the data are contradictory and 
of poor quality. We remind the reader that no Breit-Wigner resonance solution (see table 1 and figure \ref{fig:1}) is 
able to follow the trend of experimental data; the solution either reproduces the size of the 1710 MeV peak and 
under-shoots the 1780 MeV range ({\it Sol \ 2}) or the other way around ({\it Sol 1}).

As we have mentioned in the Introduction, having an "extra" P$_{11}$ state is not altogether a novelty. A second P$_{11}$ state at 1740$\pm$11 MeV is reported to be seen in the PWA analysis of ref. \cite{Bat98} in 1995, and has already been mentioned by Diakonov et. al. in ref. \cite{Dia97} as an alternative possibility to identifying the standard N(1710)~P$_{11}$ state with a $P_{11}$ cryptoexotic resonance predicted by the pentaquark models. Similarly, in ref. \cite{Arn04a}, the ordinary nucleon state has been allowed to have non-negligible admixture of a $\overline{10}$ state. Taking this mixture into account, and claiming that they have improved the standard procedures used in partial wave analysis which may miss a narrow resonance with $\Gamma \leq $ 30 MeV, the authors have estimated the width of a possible, but yet undiscovered N$^\star$ state of a mass of 1680 (1730) MeV to be $\Gamma _{N^\star \rightarrow N \pi} \approx$~2.1~(2.3)~MeV.
 \\
 \\
{\bf To conclude the chapter let us a summarize:} 

In spite of the rather bad agreement among different experiments, and the low quality of all obtained solutions given 
in table 1 (shown in figures \ref{fig:1} and \ref{fig:2}), the probability for the existence of at least one P$_{11}$ 
resonance is increased in order to explain the experimental features of the \mbox{$\pi$N $\rightarrow$ $K\Lambda$} 
process. 

The \mbox{$\pi$N $\rightarrow$ $K\Lambda$} channel offers the ideal reaction to analyze the N(1710)~P$_{11}$ resonance 
because only P-waves contribute to the process in addition to the "trivial" S-wave. Interference problems with higher 
partial waves are avoided and the angular features of the differential cross section will reflect themselves directly 
onto the properties of the N(1710)~P$_{11}$ state.\footnote{The situation resembles the $\Delta$ resonance dominance 
in the $\pi$N elastic scattering and the N(1535) S$_{11}$ importance in the $\pi$N $\eta$ production.}

We emphasize that the existing data are of extremely poor quality, even inconsistent and controversial in the W 
$\approx$ 1700 MeV range, exactly where a signal from the 1700 MeV resonant state should be visible. Therefore, in 
order to avoid model dependent amalgamation procedure, we call for a fast re-measurement of the process having in mind 
the failure of standard models to reproduce the $\sigma _{tot} (1700) \approx$ 900 nb peak. Such measurements are 
already proposed \cite{newKLambda}, but unfortunately not yet approved. 

\section{The interpretation of the P$_{11}$(1710) resonant structure}
In previous chapter we have demonstrated that within the scope of the Breit-Wigner resonance model the \mbox{$\pi$N 
$\rightarrow$ $K\Lambda$} experimental data support the existence of the P$_{11}$(1710) resonant structure. We do 
expect that such a conclusion will be confirmed within the framework of the more complex coupled-channel calculation 
revealing new quantitative information about partial wave resonant structure. 

Addressing the $qqq$ or $qqqq\bar{q}$ nature of the expected $P_{11}$ resonant structure is a tough problem because we 
have to determine what is the signal that distinguishes whether the resonant state is of three or five quark nature. 
As we may write down the Fock space expansion of any/each baryon as a linear combination of $qqq$ and $qqqq\bar{q}$ 
states, we are left with a problem of distinguishing between a three quark state with the diquark contribution coming 
from the sea quarks, and the genuine pentaquark states.
As for yet, we can not give a firm and unique definition of such a signal so we can only draw attention to some 
indications which of the configurations, three or five quark, is more probable for the established P$_{11}$(1710) 
resonant structure. Such an analysis is presently based only on the similarity of the mass, total width and branching 
ratio of the physically observed P$_{11}$(1710) resonant structure to either of the quark model predictions. We 
suggest that the decisive factor is the size of the branching ratio to inelastic strange channels, to $K\Lambda$ 
channel in our case.

As a start, we summarize the existing knowledge about characteristics of resonant states in three-quark and pentaquark 
models.

\begin{table}[!h]
\begin{center}
\caption{ Resonance parameters for the N(1710)~P$_{11}$ resonance in the three quark model. } \vspace{0.3cm}
\label{table2} 
\begin{tabular}{crcccccr}
\hline \hline
{} & {} & $\pi N$ & $\eta N$ & $K\Lambda$ & $\pi \Delta$ & \multicolumn{1}{c}{$\rho N$} & {} \\
\hline
ref. [35-38] & $\Gamma_{a} [MeV]$ & 18 & 32 & 8 & 193 & 14 & $\Gamma_{tot} = 265$\\
{} & $x_{a}$ & 0.07 & 0.13 & 0.03 & 0.73 & 0.06 & $ { x_{K\Lambda}}/{x_{\pi N}} = 0.42$\\
\hline
ref. \cite{Kon80} & $\Gamma_{a} [MeV]$ & 45 & 9 & 4 & 12 & 36 & $\Gamma_{tot} = 106 $\\
{} & $x_{a}$ & 0.42 & 0.09 & 0.04 & 0.11 & 0.34 & $ { x_{K\Lambda}}/{x_{\pi N}} = 0.09$ \\
\hline \hline
\end{tabular}
\end{center}
\end{table}

The only three quark models which are giving some predictions for the branching ratio of the N(1710)~P$_{11}$ state to 
the $K\Lambda$ channel (and are therefore relevant for our discussion) are coming from refs. \cite{Kon80} and 
\cite{Cap93,Cap94,Cap98,Cap98a}. They are summarized in table 2. \\
Only a few firm statements can be made upon observing the table.

First, model predicted total widths of the P$_{11}$ resonance are bigger then 100 MeV ($\Gamma \geq 100$ MeV).

Second, the branching ratio of the N(1710)~P$_{11}$ state to the $K\Lambda$ channel is much lower then the branching 
ratio to the $\pi N$ channel (${ x_{K\Lambda}}/{x_{\pi N}}\leq 0.42$). That is the consequence of the fact that these 
models do not have any direct mechanism of strange particles production so they both have difficulty to make the 
branching ratios to the $K\Lambda$ channel comparable to the experimental value (${ x_{K\Lambda}}/{x_{\pi N}} \approx 
1$) \cite{PDG}. Other quark models do not even give any predictions for the decay to strange channels.

On the other hand, pentaquark models predict the total width of the non-strange pentaquark to be lower then 40~MeV and 
the branching ratio to the $K\Lambda$ channel to be comparable to the branching ratio to the $\pi N$ channel 
(${x_{K\Lambda}}/{x_{\pi N}}\approx 1$).

Let us summarize: the pentaquark models predict the total width of the P$_{11}$ state to be lower, and the branching 
ratio to the $K\Lambda$ channel to be much higher then standard three quark models. And that is the only criterion we 
can offer to determine whether the physical N(1710)~P$_{11}$ state \cite{PDG} is of a $qqq$ or $qqqq\bar{q}$ nature. 
Unfortunately, the criterion is only of a qualitative, and not of a quantitative nature.

\begin{table}[!htb]
\begin{center}
\caption{ The "criteria" for discerning nature of N(1710) P11 resonance \vspace*{0.3cm}}
\label{table2a}
\begin{tabular}{cccc}
\hline \hline
{} & {} & {} & {}\\ [-1.5ex]
{} & {$\Gamma[MeV]$} & {$\sqrt{x_{\pi N} x_{K\Lambda} }$} & {${x_{K\Lambda}}/{x_{\pi N}}$}\\[+1ex] 
\hline \hline
{}& {}& {}& {}\\[-1.5ex]
$PDG$ & 100 & 0.15 & 1.00 \\ 
$Sol \ 1$ & 180 & 0.22 & 1.05 \\ 
$Sol \ 2$ & 180 & 0.28 & 1.59 \\ 
$Sol \ 3$ & {\bf 68} & 0.26 & 1.45 \\ \\ [-2.ex]
$Sol \ 4$ & {\bf 68} & 0.26 & 1.45 \\ [-0.5ex]
 & {\bf 154} & 0.08 & 0.09 \\ 
\hline \hline
\end{tabular}
\end{center} 
\end{table}

It is interesting to observe that the PDG values for the N(1710)~P$_{11}$ state do not exactly follow the suggested 
"criteria" neither for the three nor for the pentaquark models (as seen in table 3). They partly follow the trend of 
the three quark models for the width, but the trend of the pentaquark model for the branching ratio. The width of the 
N(1710)~P$_{11}$ state is rather undetermined but bigger the 100 MeV ($90<\bar{\Gamma}_{\rm PDG}< 480$~MeV), while the 
branching fraction to the $K\Lambda$ channel is not low, but quite comparable to the branching fraction to the $\pi N$ 
channel. 

The afore discussed criteria are summarized in table \ref{table2a}. The criteria are indecisive for $Sol \ 1$ and $2$, 
and are qualitatively the same as for PDG results. For $Sol \ 2$ the branching ratio to $K\Lambda$ channel gets 
bigger, but the partial width is so big that the resonant states should be of three quark origin. The $Sol \ 3$ and 
$Sol \ 4$ are definitely closer to what would one expect that the pentaquark state should look like. The branching 
fraction to the $K\Lambda$ channel exceeds the branching fraction of the $\pi N$ channel ( ${ x_{K\Lambda}}/{x_{\pi 
N}}\approx 1.45$), and the width tends to go below 90 MeV. We believe that the {\it Sol 3} is not acceptable because 
of problems of the total cross section in the 1750 MeV range, and to our "taste" the $Sol \ 4$ looks very much like a 
mixture of a pentaquark (low $\Gamma$ and high ${x_{K\Lambda}}/{x_{\pi N}}$), and a standard three quark state (high 
$\Gamma$ and low ${x_{K\Lambda}}/{x_{\pi N}}$). However, we can not offer any proof that one of the two suggested 
states indeed is a pentaquark. 

Accepting the arguments of ref. \cite{Pra04,Pak04} that it is impossible to produce the N(1710)~P$_{11}$ state as a 
pure pentaquark state, one of the solutions to the problem is that the physically observed P$_{11}$ state is a mixture 
of two states: one of three-quark and second of pentaquark nature. However, the nature may complicate matters even 
more because the hybrid baryons of the type $(qqqG)$ may easily contribute to the P$_{11}$ wave as well, so this state 
can be a simultaneous admixture of both, pentaquarks and hybrid baryons to the $(qqq)$ state. The whole problem of 
hybrid $(qqqG)$ barons is extensively discussed in literature. We refer the reader to the summary presentation of the 
topic given in ref. \cite{qqqG}. 
\section{Conclusions}
The Breit-Wigner resonance model calculation using the standard set of parameters for the N(1650)~S$_{11}$, 
N(1710)~P$_{11}$ and N(1720)~P$_{13}$ resonances of ref. \cite{PDG} is not able to understand the $\sigma _{tot} 
\approx$ 900 nb peak. It reproduces the overall absolute value of the total cross section reasonably well, and manages 
to reproduce the shape of the angular distribution only at W=1683 MeV. However, it fails miserably in reproducing the 
shape of the differential cross section at other energies.

The new Breit-Wigner resonance model fit with only one P$_{11}$ resonance has improved the shape of the differential 
cross sections at all energies, and indicates that the contribution of the N(1720)~P$_{13}$ resonance is negligible. 
The 900 nb total cross section peak is not yet understood. 

The faithful reproduction of the 900 nb total cross section peak of the amalgamated total cross section data can be 
achieved only if the N(1710)~P$_{11}$ resonance is fairly narrow ($\Gamma \approx$ 68~MeV), but then the agreement 
with the total cross section data at higher energies is spoiled. 

The simultaneous reproduction of the $\sigma _{tot} \approx$ 900 nb peak and the higher energy total cross section 
values can within the Breit-Wigner resonance model be only achieved if the existence of another P$_{11}$ resonance is 
assumed. 

Addressing the $qqq$ or $qqqq\bar{q}$ nature of the resonant structure of total cross section in the energy range of 
1700~MeV we re-open the possibility that the additional P$_{11}$ resonance present in 1700~MeV energy range in 
addition to the standard N(1710)~P$_{11}$ PDG-resonance \cite{PDG} is a cryptoexotic pentaquark state. The second 
P$_{11}$ state is the prediction of the Zagreb group made in 1995. \cite{Bat98}, and already mentioned as a possible 
physical realization of the pentaquark model in the original Diakonov et. al. publication \cite{Dia97}.

In order to confirm that the N(1710)~P$_{11}$ resonant state is, or at least has the admixture of a non-strange 
pentaquark, the proper distinguishing criteria between $qqq$ or $qqqq\bar{q}$ have to be developed. The present 
qualitative considerations are only indicative.

The precise measurement of the total cross section and angular distributions of the $\pi$N $\rightarrow$ $K\Lambda$ 
process in the energy range \mbox{1613 MeV $<$ W $<$ 1900 MeV} is urgently needed and strongly recommended, because no 
serious theoretical analysis can be performed on the basis of present data set.

The decisive conclusion about the existence of the $P_{11}$ non-strange pentaquark will be possible only when the 
improved set of data is fully incorporated in one of the existing coupled-channel partial wave analyses 
\cite{Bat98,Arn85,Arn95,Arn04,Cut79,Vra00,Man84,Man92,Hoh83,Gre99} using all, elastic and inelastic experimental data 
in the 1650-1850 MeV energy range. Only then the minimal necessary number of P$_{11}$ resonances will be unambiguously 
determined.
The elimination of a cusp effect as a source of the observed structure will be as well achieved in the framework of a 
coupled-channel, multi-resonance theoretical analysis. As the the $K \Sigma$ channel is the most probable reason for 
the cusp effect it has to be included in such an analysis. The data for the $\pi$N $\rightarrow$ K$ \Sigma$ process in 
the energy range \mbox{1683 MeV $<$ W $<$ 1900 MeV} are scarce \cite{KSigma} so the measurement of the total cross 
section and angular distributions for that process are badly needed as well.
In this article, using the Breit-Wigner resonance model only, we show a strong indication that a standard, wide 
P$_{11}$ resonance is incompatible with the existing data, and that a narrow P$_{11}$ state weakly coupled to the $K 
\Sigma$ channel is quite probable.

\subsection*{Acknowledgement}
We are grateful to K. Kadija from Rudjer Bo\v{s}kovi\'{c} Institute, a member of Na49 Collaboration, for a constant 
encouragement to link our "P$_{11}$" problems with the new pentaquark states. We also thank M. Korolija from the same 
Institute for initiating out interest in $\pi$N $\rightarrow$ $K\Lambda$ process.


\begin{thebibliography}{99}
\bibitem{Kle03} For the extensive experimental review see: E. Klempt, Preprint \mbox{hep-ph/0404270v1}.
\bibitem{Dia97} D. Diakonov, V. Petrov and M. Polyakov, Z. Physik {\bf A359}, 305 (1997).
\bibitem{Dia03} D. Diakonov and V. Petrov, Phys.Rev. {\bf D69}, 09411 (2004).
\bibitem{Jaf03} R. Jaffe and F. Wilczek, Phys.Rev.Lett. {\bf 91}, 232003 (2003).
\bibitem{PDG} Particle Data Group: Review of Particle Physics; K. Hagiwara et.al., Phys.Rev. {\bf D66}, 010001 (2002).
\bibitem{Bat98} M. Batini\'{c}, I. \v{S}laus, A. \v{S}varc and B.M.K. Nefkens, Phys.Rev. {\bf C51}, 2310 (1995);
M. Batini\'{c}, I. \v{S}laus, A. \v{S}varc, B.M.K. Nefkens and T.S.-H. Lee, Physica Scripta {\bf 58}, 15(1998).
\bibitem{Pra04} M. Praszalowicz, Annalen der Physik {\bf 13}, 709 (2004).
\bibitem{Pak04} S. Pakvasa and M. Suzuki, Phys.Rev. {\bf D70}, 036002 (2004).
\bibitem{Lan99} L.G. Landsberg, Physics Reports {\bf 320}, 223 (1999).
\bibitem{Wen04} Wen-Tai Chiang, B. Shaghai, F. Tabakin and T.S.-H. Lee, Phys. Rev. {\bf C69}, 065208 (2004).
\bibitem{Pen02} G. Penner and U. Mosel, Phys.Rev. {\bf C66}, 055211 (2002).
\bibitem{Shk05} V. Shklyar, H. Lenske and U. Mosel, Phys.Rev {\bf C72}, 015210 (2005).
\bibitem{Hoh05} G. H\"{o}hler, private communication and A. \v{S}varc, Invited talk at The 2nd International
Pion-Nucleon PWA Workshop, Zagreb 27.06 - 03.07.2005.
\bibitem{Arn85} R.A. Arndt, J.M. Ford and L.D. Roper, Phys.Rev. {\bf D32}, 1085 (1985).
\bibitem{Arn95} R.A. Arndt, I.I. Strakovsky, R.L. Workman and M.M. Pavan, Phys.Rev. {\bf C52}, 2120 (1995).
\bibitem{Arn04}R. A. Arndt et. al., Phys.Rev. {\bf C69}, 035213 (2004).
\bibitem{Arn04a} R. A. Arndt, Ya. I. Azimov,M- V. Polyakov, I. I. Strakovsky and R. L. Workman, Phys.Rev. {\bf C69}, 
035208 (2004).
\bibitem{Hoh83} G. H\"{o}hler, Elastic and Charge Exchange Scattering of Elementary Particles, Vol. {\bf 9}, Subvolume 
{\bf b/II}: Pion Nucleon Scattering, Landolt-B\"{o}rnstein, (1983).
\bibitem{Cut79} R.E. Cutkosky, C.P. Forsyth, R.E. Hendrick and R.L. Kelly, Phys.Rev. {\bf D20}, 2839 (1979).
\bibitem{Vra00} T.P. Vrana, S.A. Dytman and T.S.-H- Lee, Physics Reports {\bf 328}, 181 (2000).
\bibitem{Feu98} T. Feuster and U. Mosel, Phys.Rev. {\bf C58}, 457 (1998).
\bibitem{Cut90} R.E. Cutkosky and S. Wang, Phys.Rev. {\bf d42}, 235 (1990).
\bibitem{Hoh93} G. H\"{o}hler, {\it $\pi$N Newsletters} {\bf 9}, 1 (1993).
\bibitem{Bak77}R.D. Baker et. al., Nucl. Phys. {\bf B126}, 365 (1977).
\bibitem{Bak78} R.D. Baker et. al., Nucl.Phys. {\bf B141}, 29 (1978).
\bibitem{Dyc69} O.Van Dyck, et al., Phys.Rev.Lett. {\bf 26}, 860 (1971).
\bibitem{Jon71} J.J. Jones et al., Phys.Rev.Lett. {\bf 23}, 50 (1969).
\bibitem{Kna75} T.M. Knasel et al., Phys.Rev. {\bf D11}, 1 (1975).
\bibitem{Loc86} M.P. Locher, M.E. Sainio and A. \v{S}varc, Adv.Nucl.Phys. {\bf 17}, 47 (1986).
\bibitem{Man84} D.M. Manley, R.A. Arndt., Y. Goradia and V.L. Teplitz, Phys.Rev. {\bf D30}, 904 (1984).
\bibitem{Man92} D.M. Manley and E.M. Saleski, Phys.Rev. {\bf D45}, 4002 (1992).
\bibitem{Gre99} A.M. Green and S. Wycech, Phys.Rev. {\bf C60}, 035208 (1999).
\bibitem{newKLambda} D. Isenhower, et al, Proposal to Upgrade the MIPP Data Acquisition System at Fermilab, March 
2005, I.G. Alekseev et al, Search for the Cryptoexotic Member $N_{\bar{10}}$ of the Baryon Antidecuplet $1/2 ^+$ in 
the Reactions $\pi ^- p \rightarrow \pi ^- p$ and $\pi ^- p \rightarrow K\Lambda$, Experiment Proposal from ITEP-PNPI 
Collaboration, May 2005.
\bibitem{Kon80} R. Koniuk and N. Isgur, Phys.Rev. {\bf D21}, 1868 (1980).
\bibitem{Cap93} S. Capstick and W. Roberts, Phys.Rev. {\bf D47}, 1994 (1993).
\bibitem{Cap94} S. Capstick and W. Roberts, Phys.Rev. {\bf D49}, 4570 (1994).
\bibitem{Cap98} S. Capstick and W. Roberts, Phys.Rev. {\bf D57}, 4301 (1998).
\bibitem{Cap98a} S. Capstick and W. Roberts, Phys.Rev. {\bf D58}, 074011 (1998).
\bibitem{KSigma} Landolt-B\"{o}rnstein, New Series, ed. H. Schopper, {\bf 8} (1973).
\bibitem{qqqG} P. R. Page, Invited plenary talk presented at the 10th International Symposium on Meson-Nucleon Physics 
and the Structure of the Nucleon (MENU 2004), Beijing, Int.J.Mod.Phys. {\bf A20}, 1791 (2005).
\end{thebibliography}
\end{document}